\documentclass[12 pt]{article}
\usepackage{graphics}
\usepackage[dvips]{epsfig}
\usepackage{amsmath}
\usepackage{xcolor}

\begin{document}
	
\title{A machine learning based control of chaotic systems}
\author{	P. Garc\'{\i}a\\
\small		Laboratorio de Sistemas Complejos,\\
\small		Departamento de F\'{\i}sica Aplicada,\\
\small		Facultad de Ingenier\'{\i}a, 
\small		Universidad Central de Venezuela.\\
		
\small		and\\
		
\small		Red Iberoamericana de Investigadores en
\small		Matemáticas Aplicadas a  Datos. AUIP.
}

\date{}
	
\maketitle
\thispagestyle{empty}

\begin{abstract}
	In this work, inspired by symbolic dynamic of chaotic systems and using
	machine learning techniques, a control strategy for complex systems is designed.
	Unlike the usual methodologies based on modeling, where the control signal is
	obtained from an approximation of the dynamical rule, here the strategy rest
	upon an approach of a function that, from the current state of the system, give
	the necessary perturbation to bring the system closer to a homoclinic orbit that
	naturally goes to the target.
	The proposed methodology is data-driven or can be developed in a 
	model-based context and is illustrated with computer simulations of chaotic systems
	given by discrete maps, ordinary differential equations and coupled map networks.
	Results show the usefulness of the design of nonlinear control techniques based on
	machine learning and numerical approach of homoclinic orbits.
\end{abstract}

\section{Introduction}
\setcounter{equation}{0}
There are  experimental and/or numerical evidence of the chaotic behavior in systems from:  population dynamics\cite{Din}, epidemiology\cite{Upadhyay}, ecosystems\cite{Singh}, cardiac rhythm\cite{Zhao}, neurological systems\cite{Hu}, optoelectronic systems\cite{Ghosh}, chemical reactions\cite{Xu},  economic systems\cite{Alves},  communication systems\cite{Mukherjee} and a lot of more situations  in physics, chemistry, biology and  engineering\cite{Elhadj}. Added to this, that behavior can be useful in areas such as secure communication, where chaotic systems offer an alternative for the development of new secure communication technologies \cite{Rosier, Teh}.
	
As can be seen in most of the before mentioned cases, it is desirable to have strategies that allow the evolution of the system to be regulated: the population of animal communities can be controlled to prevent extinctions or the spread of diseases, for obvious reasons the epidemic diseases must be controlled, the heart rate must be controlled in the case of arrhythmia, brain activity must be controlled in case of epilepsy or finally some chemical, physical or computational systems can be controlled for the benefit of all. This represents sufficient motivation to devote effort to the design models to forecast or control schemes for such systems.
	
The origin of chaotic behavior in dynamic systems can be explained, at least partially, from two different perspectives: the first is quantitative or statistical and is associated to the determination of quantities such as Lyapunov exponents that discriminate between chaotic or non-chaotic behavior. The second approach involves a geometric view for the dynamical system and attempts to isolate the geometrical structures that support the chaotic motion. The basic geometric objects, in this case,  are the stable and unstable manifolds of fixed points of maps, or critical points or periodic orbits of ordinary differential equations. The homoclinic orbits are orbits that lies in both the stable and unstable manifolds of fixed points, or periodic orbits and particularly joins a saddle equilibrium point to itself. 
	
The effect of the intersection of the stable and unstable manifolds, i. e., the homoclinic orbit, is nevertheless felt globally and offers a way in which simple local information can be extrapolated to complicated global behavior \cite{Jones}.

From a general perspective a dynamical system can be modeled as $ s_t = f_{r}^t (s_0) $, where $ r $ is a set of parameters, $ t \in \Re $ or $ t \in \cal N $ represents the time, $ s \in S $ the state of the system and $ f $, a function that regulates the evolution of the system; so that, the control over them can only be achieved applying disturbances of some of these components. \\
	
When systems have chaotic behavior: on the one hand they have sensitivity to the initial conditions, which makes them difficult to predict in the long term, but on the other it offers the opportunity to change their behavior radically using small perturbations, this is, with a low cost and without producing important alterations in the system.\\ 
The seminal article on this topic presents a method for the control of chaotic systems, known as OGY method\cite{Ott} and where the term {\it controlling chaos } was coined, is the archetype of the methods that belong to the class of those that disturb the parameter. As evidence of its importance, from the point of view of the basic research or the technology of the finding of Ott, Grebogi and York, there is the great amount of methods to stabilize chaotic systems that have been designed later, see example \cite{Souza} and references therein.
Another method almost as celebrated, as the previous one, is a feedback method proposed by Pyragas {\it et. al.} \cite{Pyragas} belonging to the methods that disturb $ f $. In this method, the stabilization of unstable periodic orbits of a chaotic system is achieved by combined feedback with the use of a specially designed external oscillator. Neither of the two methods require an {\it a priory} analytical knowledge of the dynamics system and may be applicable in experimental situations. 
	
Among the control strategies of chaotic systems,  there are a significant amount of control methods based on linear approximations of the dynamics around the target.  When is it so, it is necessary that the linear approach is a good representation of the system or equivalently that the trajectory of the system is close enough to the target, which makes the control less efficient.
	
In this work, unlike those that model the dynamics in linear or nonlinear form, to design the control function, here  a nonlinear control function is directly obtained by using Kernel Regression methods\cite{Schlkopf} and data from the observation of the system. The training set for the kernel machine, in this case, is constructed using a set of states that are close to points of a homoclinic orbit.
	
	The article is organized as follow: in Section 2 a symbolic dynamic inspired but  machine learning based approach to control of chaotic dynamic is presented. In Section 3, two versions of the solution to the problem are presented, depending on whether or not the symbolic dynamics of the system are known. Section 4 is devoted to give out some examples of the performance of the methodology and finally in Section 5 we give some concluding remarks.
	
	\section{A machine learning approach to control chaos}Colloquially, a control problem can be paraphrased as a sequence of three steps: (i) the identification of the control's object, (ii) the selection of the control's target and (iii) the design of a control strategy.
	
	\subsection{Control object and target}
	We will consider chaotic systems in one or more dimensions, continuous or discrete and represented by discrete maps, ordinary differential equations or coupled map networks.  However, in order to show the ideas behind our control problem in a simple form, we will start considering as control object a chaotic, one-dimensional and uni-modal map $ s_ {t + 1} = f_ {r} (s_t) $,  for later in the results section use as control objects the rest of previously listed systems.\\
	The targets, in all examples, are given by a unstable fixed-point solutions (${\bf s}^*$) of the dynamical system, although the control strategy can be adapted to the case of other types of periodic orbits.
	
	\subsection{Control strategy}
	Our approach to the control strategy propose to  the estimate a small perturbation $u_t$, to the actual state $s_t$, in such a way that in the $(t+1)$-th iteration of $f$ the orbit come close to some point  $s_0$, whose evolution has the form $s_0, s_1, \cdots, s_{k-1}, s^*+\delta$, where $\delta $ is a small value. This is, we should to estimate a perturbation  $u_t$, such that, if
	\begin{equation}
	s_{t+1} = f_r(s_t)+u_t,
	\end{equation}
	
	\noindent
	then $|s_{t+1}-s_0|<|s_t-s_0|$, with $s^* + \delta = f^{k}_r(s_0)$. 
	
	In some sense, the idea behind this methodology is like to the OGY method, it consists in bringing the system closer, not immediately, to the control's target but to  points of orbits that naturally takes it as close as possible to target. However, unlike the OGY method, we do not disturb the parameters of the system or use explicitly a linear approximation of the system.
	
	Thus, if we are able to estimate these new points and include it in the target, we expect that the control scheme to be more effective. The points that fulfill that condition, in the case of the fixed points of chaotic systems, are the points of {\it homoclinic orbits}. This orbits joins a saddle equilibrium point to itself\cite{Block}, and offer a number of potential targets equal to the points in this orbit. 
	
	So, our control strategy start by determining the set of estates $\{ s^i_0 \}_{i=1}^q$. Clearly, these states  can be estimated if we iterate the inverse function of $ f $ (evolving in time reversal the system),  from $ s ^ * + \delta$, this is,  $s_0 = f^{-k}_r(s^* + \delta)$, but this can no be done simply because the inverse of a nonlinear function is multi-valued. 
	
	However, it is well known\cite{Hao} that the phase space evolution of an uni-modal discrete map can be translated, in bi univocal way, into a binary representation by using a partition of the state space in two regions, labeled each of them by one symbol ($\sigma_t$), $0$ or $1$. The symbolic representation is obtained replacing $s_t $ by the label associated to the element of the partition which belongs $ s_t \to \sigma_t $. With this new information $f_r^{-1}$, can be written as as:
	\begin{equation}
	f_r^{-1}(s_t,\Sigma) = \left\{
	\begin{array}{ccc}
	f_{r,0}^{-1}(s_t) & if  &\sigma_t = 0 \\
	f_{r,1}^{-1}(s_t) & if  &\sigma_t = 1 \\
	\end{array} \right.,
	\label{s0}
	\end{equation}
	
	\noindent
	where $ f_{r, 0} $ and $ f_{r, 1} $ are the monotone branch at the left or right from the maximum of $ f $, respectively. In this form, every state $s_t$, can be represented by a sequence of symbols $ \Sigma = \{ \sigma_{j}\}_{j=1}^{l} $, given by the symbolic chain generated from the initial condition $ s_0 $. Here $l$ give the accuracy with which $s_0$ is represented.  In particular, the states from which the system reaches a neighborhood of the fixed point are given by:
	
	\begin{equation}
	s_0 = f_{r,\sigma_1}^{-1} \circ \cdots  \circ  f_{r,\sigma_{l-1}}^{-1}  \circ   f_{r,\sigma_l}^{-1} (s^*+\delta) 
	\label{it-inv}
	\end{equation}
	
	For $l$ large  enough,  the iteration of (\ref{it-inv})  converges approximately to the real number $s_0$ independently of the value of $\delta$,  if the sequence of symbols $\sigma_t$ is appropriate, i. e. if these  sequences are of the form
	\begin{equation}
	\Sigma_i = \sigma^i_{1},\sigma^i_{2}, \cdots, \sigma^i_{k}, \underbrace{\sigma^*, \cdots, \sigma^*}_{(l-k)~times}
	\label{homoclinic}
	\end{equation}
	
	\noindent
	are large enough and $\sigma^*$ is the symbol associated to the interval that $s^*$ belongs. 
	
	The sequences $\Sigma_i$ form a set of $2^k$ different binary chains.  Not all of them are {\it admissible} (see for example \cite{Hao}, Sec. 2.5.5), but the existence of some of them is guaranteed by the presence of homoclinic orbits in the attractor of $f$. This sequences can be used as a symbolic representation of segments of homoclinic orbits and will be used to estimate the $ 2^k $ initial conditions, 
	
	\begin{equation}
	s^i_0 = f_r^{-l} (s_{R_i}, \Sigma_i) 
	\end{equation} 
	
	\noindent
	with $s_{R_i}$ a random real number in the adequate domain, which places the system in a orbit towards the target $ s^* $.

	Now, in order to estimate the $u_t$ we start generating, from $q$ strings like (\ref{homoclinic}),  $q$ initial conditions $S_0 = \{ s^1_0, s^2_0, \cdots, s^q_0 \} $, from which the system reach a neighborhood of $s^*$ in $l$ iterations. 
	
	\begin{figure}[h]
		\centerline{
			\includegraphics[width=0.7\textwidth]{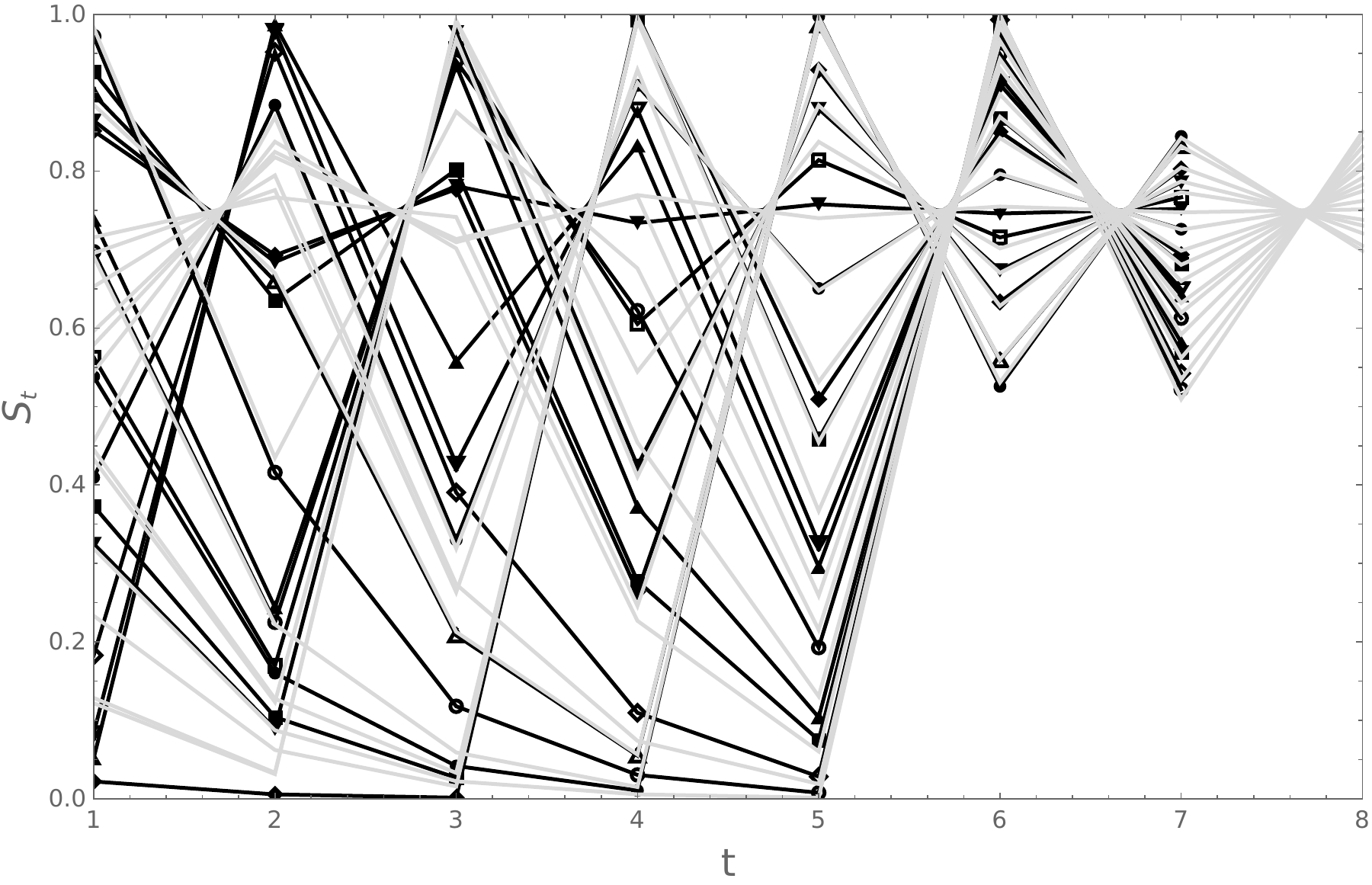}
		}
		\caption{\small Initial conditions $s^i_0 $ and its trajectories, obtained from $\Sigma_i $. The black lines shows the evolution of $16$ initial conditions, with $l=7$ and $k=4$. The gray ones, the evolution of the same number of initial equations with $l=8$ and $k=4$. }
		\label{Fig0}
	\end{figure}
	
	With the before estimate states, the $N$ pairs $ \{u_t, s_t\}_{t=1}^{N} $ are generated using,
	
	\begin{equation}
	u_t = P(S_0,s_{t}) - s_{t},
	\end{equation}
	
	\noindent
	from the elements of an typical orbit $ \{s_t \} _ {t = 1}^N $ of the system. Here, $P(S_0,s_{t})$ is a function that pick out,  the element of $S_0$ nearest to $s_ {t}$ is defined as:
	
	\begin{equation}
	P(S_0,s_t) = argmin_{s^i_0}{\{|s^i_0-s_{t}|\}}.
	\end{equation}
	
	\noindent
	These pairs will later serve as training data for a nonlinear regression scheme $ u_t= u(s_{t}) $. 
	In this way the control problem is turned into a regression problem, that we propose to solve using machine learning techniques.
	
	As one, among many alternatives to approximate $u$, in this work we use a powerful set of strategies from machine learning known as Kernel Regression Methods \cite{Schlkopf, Garcia}. 
	
	The main idea behind of these techniques is to map the observed examples (training set) into a new representation space (known as feature space), where object properties can be extracted more easily, using  linear algorithms.
	
	With the aim of showing how one of these techniques can be selected according to the particularities of the problem, in this case we present results with Gaussian Process\cite{Rasmussen} in the case of problems associated with a small number of data and Random Forest\cite{Breiman} in the case that the problem involve a large amount of data.
	
	\subsubsection{Gaussian Process Regression}
	Gaussian processes is a kernel-based Bayesian tool for modeling problems. Like other kernel-based methods, it combine a high flexibility of the model by working in often infinite-dimensional feature spaces with the simplicity that all operations are performed in the lower-dimensional
	input space using positive definite kernels. 
	
	Here we will understand these processes  according to their usual definition:  a set of indexed random variables,  such that every finite subset of those random variables has a multivariate normal distribution. The distribution of a $GP$ is the joint distribution of all those random variables, and as such, it is a distribution over functions with a continuous domain.
	
	These  process are specified by its {\it mean function} $m(s)$ and {\it covariance function} $k(s, s_i)$ and are a natural generalization of the Gaussian distribution whose mean and covariance are represented by a vector and a matrix, respectively.  We will represent this process, with the usual notation, as:
	\begin{equation}
	u \sim GP(m(s), k(s,s')),
	\end{equation}
	
	\noindent
	and we will read this, as: {\it the function $u$ is distributed as a Gaussian process with  prior mean function $m(x)$ and covariance function $k(s,s')$ }. If we generate a random  vector from this distribution, $ u \sim {\cal N}(m(s)$, $k(s,s'))$,  this vector will have the function values $u(s)$ as coordinates indexed by $s$.
	
	This Gaussian process directly captures the model uncertainty but are computationally expensive in the case of great amount of data.

	\subsubsection{Random Forest Regression}
	Decision tree learning is one of the predictive modeling approaches used in statistics, data mining and machine learning. It uses a decision tree, as a predictive model, to go from observations about an item (represented in the branches) to response about the item's target value (represented in the leaves).
	
	Random forests or random decision forests are an ensemble randomized decision trees designed for classification or regression. This strategy can be seen as a kernel method \cite{Scornet, Biau} that operate by constructing a multitude of decision trees at training time and average prediction of the individual trees. 
	
	This methodology runs efficiently on large databases and reduces overfitting in decision trees, helping us to improve the accuracy.
	
	\section{Model-based and model-free versions of the method}
	
	To illustrate the versatility of the idea, we will show an implementation of two versions of it in a simple case: the first that we will call model-based, refers to the case in which the functional form of the dynamic system and its symbolic dynamics are known, the second , which we will refer to as model-free, represents the case in which only data, resulting from the observation of the system, is available.
	
	\subsection{Model-based control}
	To illustrate our strategy, let us start considering the Logistic map, $f(s_t) = 4 s_t (1-s_t)$. It has an unstable fixed point at $s^*=3/4$ that we will consider as the target. The inverse function of $f$ can be written using the symbolic dynamic:
	\begin{equation}
	f_{r}^{-1}(s_t,\Sigma) = \left\{
	\begin{array}{cc}
	\frac{1}{2} + \frac{\sqrt{1-s_t}}{2} & if ~ \sigma_t = 0 \\
	\frac{1}{2 } - \frac{\sqrt{1-s_t}}{2} &  if ~\sigma_t = 1 \\
	\end{array} \right.,
	\end{equation}
	
	The controlled system can be written as $s_{t+1} = 4 s_t (1-s_t) + u_t$, with $u_t=u(x_t)$. 
	In this case, since the symbolic dynamic is known, the model-based and data-driven control schemes are showed. Figures 1 and 2, shows  the results of the control scheme application using $500$ data points with $l=6$, $k=4$ and $q=16$ in the case of model-based control and  $l=6$ and $q=16$ in the case of data-driven control. 
	
	\begin{figure}[h]
		\centerline{
			\includegraphics[width=0.5\textwidth]{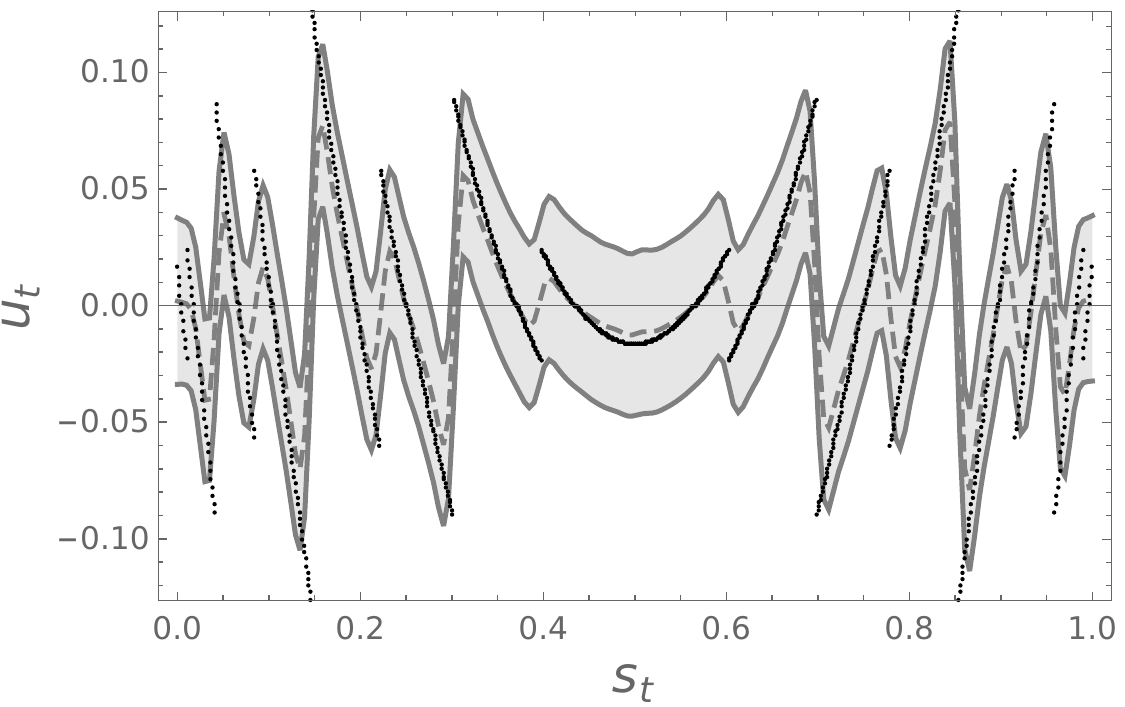}
			\includegraphics[width=0.5\textwidth]{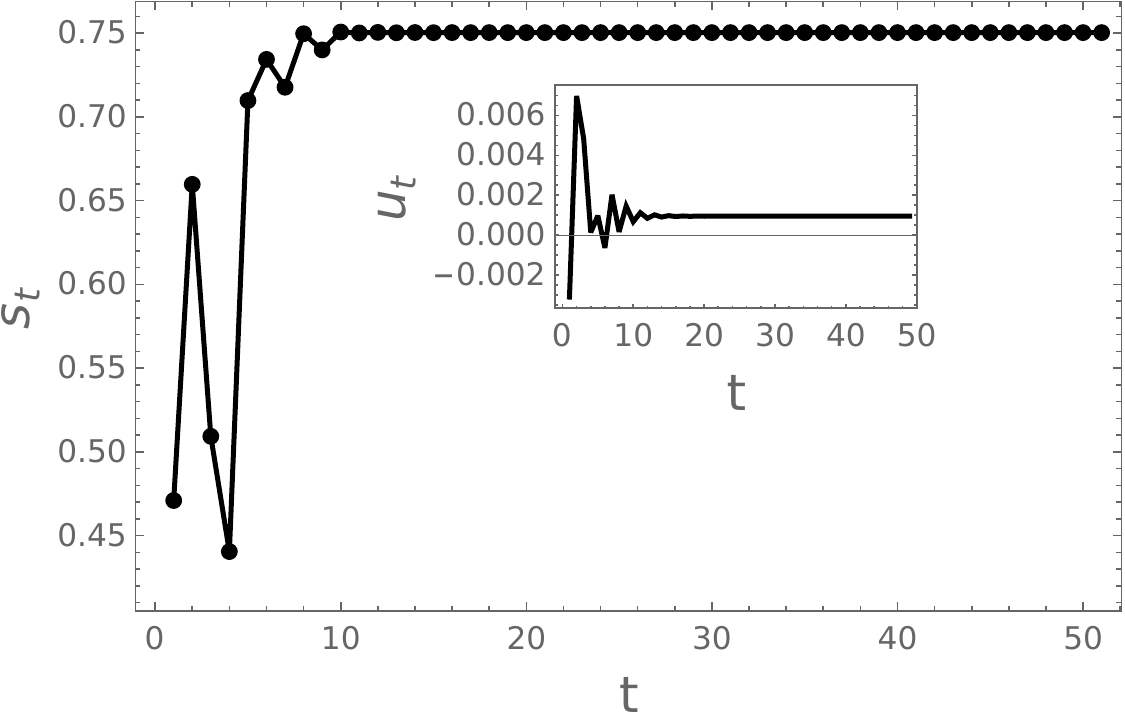}
		}
		\caption{\small Model-based control approach of Logistic map. Left: Black points represents the training set of the Gaussian process. Gray dashed line represents the predicted values of the function $u(s_t)$ and gray shadow represents the confidence interval of the predictions, given by two times the standard deviation of the errors. Right: Evolution of the controlled system. The inset shows the applied perturbations. In both cases, the results shown were averaged for $100$ initial random conditions.}
		\label{Fig1}
	\end{figure}
	
	It is worth to note that in general,  if  all possible sequences of symbolic states are considered as training data generation,  the quality of the performance of the strategy could be affected, because this number of pairs does not necessarily coincide with the admissible ones. Now, if the  dynamics are available, it is possible to determine the admissible symbolic sequences using the strategy proposed in \cite{Hao}. The study of this aspect of the problem is not the subject of this article, but will be addressed in future works.
	
	In this work, the strategy of control depend on the methodology to approximate $ u $ but we believe that any strategy, to interpolate the  response of the system to states that are not present  in the their training set, can solve the control problem.
	
	\subsection{Model-free control}
	Since it is not always possible to have the symbolic dynamics of the system, in the following we shows how it is possible to implement our idea using data from a typical orbit of the system.  Thus a data-driven version of the strategy is easily designed if we replace the symbolic dynamics knowledge with an algorithm that identifies,  in the time series,  a set of sequences of states converging to $ s^* $. 
	
	At this point, we could use a numerical method of detection of homoclinic orbits as in references \cite{Avrutin, Beyn, Korostyshevskiy} and find the points which we will include in the target. However, we will opt for a simpler strategy: we start identifying, in the time series, the set of the nearest states to $s^*$, $\{s_{m_1}, s_{m_2}, \dots, s_{m_q}\}$, so that $s^i_0 = s_{m_i-l}$, with $i=1, \cdots, q$ and $l$ is a integer identifying the $l$-th predecessor of  $s_{m_i}$. Thus the set $S_0 = \{ s_0^i\}_{i=1}^q$  is constructed and given that set  the control strategy is the same that in previous section.
	\vspace{1 cm}
	
	\begin{figure}[h]
		\centerline{
			\includegraphics[width=0.5\textwidth]{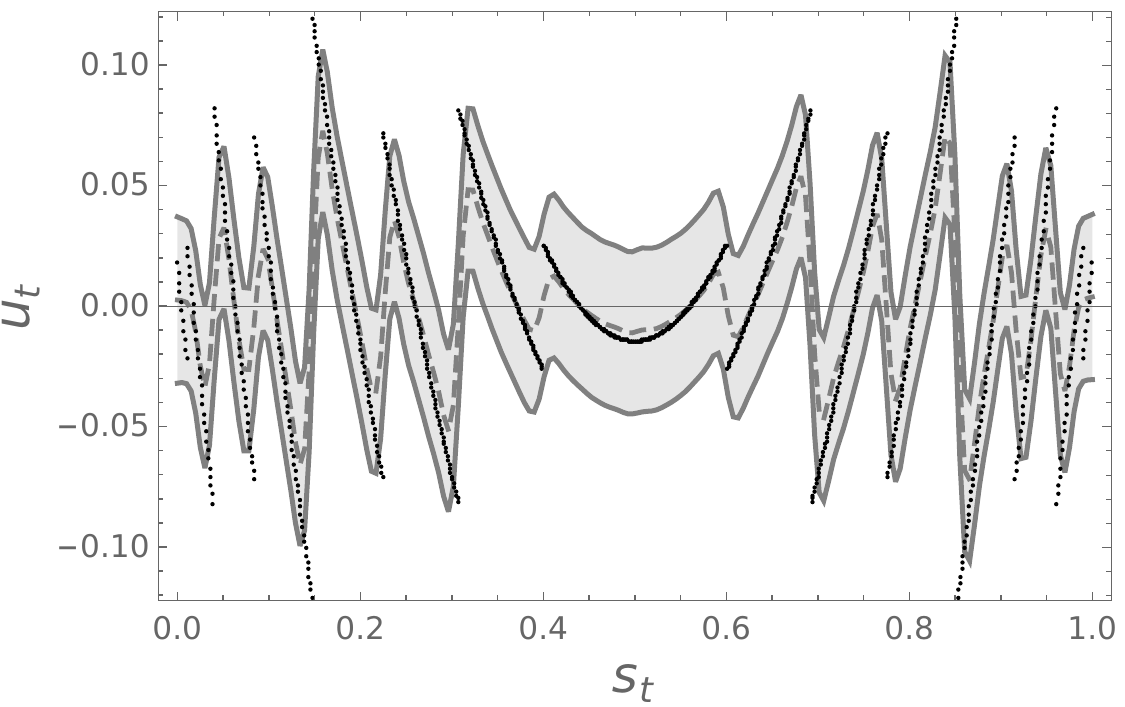}
			\includegraphics[width=0.5\textwidth]{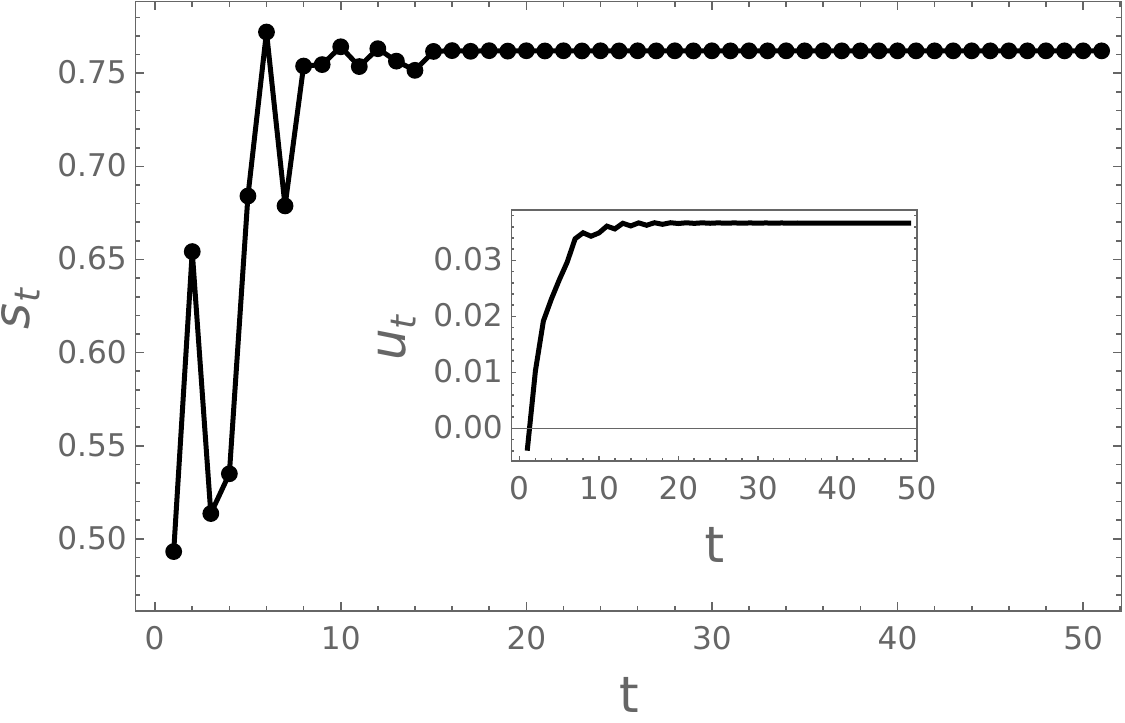}
		}
		\caption{\small Model-free control approach of Logistic map.  Left: Black points represents the training set of the Gaussian process. Gray dashed line represents the predicted values of the function $u(s_t)$ and gray shadow represents the confidence interval of the predictions, given as in the model-based case. Right: Evolution of the controlled system. The inset shows the applied perturbations. In both cases, the results shown were averaged for $100$ initial random conditions.}
		\label{Fig2}
	\end{figure}
	
	In the following section the performance of the methodology is shown. As in the examples that follow it is not so simple to obtain the symbolic representation, we will only show results of the model-free approach.
	
	The scheme can be summarized as the following list of tasks:
	
	\begin{enumerate}
		\item Give a time series $\{ s_i \}$, chose: $\sigma^*$, $q$, $k$ and	$l$
		\item Identify $q$ states $S_0 = \{ s^1_0, s^2_0, \cdots, s^q_0 \} $,
		such that,  with $s^* + \delta = f^{k}_r(s_0)$.
		\begin{itemize}
			\item If the symbolic dynamic is known:
			\begin{itemize}
				\item[i.] Identify a set of admissible states $\Sigma_i = \sigma^i_{1},\sigma^i_{2}, \cdots, \sigma^i_{k}, \underbrace{\sigma^*, \cdots, \sigma^*}_{(l-k)~times}$,
				\item[ii.] $ s^i_0 = f_r^{-l} (s_{R_i}, \Sigma_i)$ with $s_{R_i}$ a random real number.
			\end{itemize}
			\item Else:
			\begin{itemize}
				\item[i.] Identify a set of the nearest states to $s^*$, $\{s_{m_1}, s_{m_2}, \dots, s_{m_q}\}$,
				\item[ii.]  $s^i_0 = s_{m_i-l}$.
			\end{itemize}				
		\end{itemize}
		\item Construct the training set $ \{(s_t, u_t)_i\}_{i=1}^{N} $, with $u_t = P(S_0,s_{t}) - s_{t}$.
		\item Train some Machine Learning Regression scheme, to obtain, $u_t = u(s_{t})$.
	\end{enumerate}
	
	\section{Numerical Results}
	In order to show the performance of the proposed scheme, in a more general context, we use as examples: the Henon map, the Lorenz system and a network of coupled discrete maps. These represent two, three and $n$-dimensional systems with discrete time and continuous time. Each of these systems has a homoclinic orbit, as can be seen in \cite{Shi, Leonov}
	
	\subsection{Henon map}
	The Henon map is used as canonical example of bi-dimensional chaotic system and given by a difference equation system. It has an unstable fixed point at $(x^*,y^*)=(0.6313, 0.1894)$ and the controlled system is given by:
	\begin{eqnarray}
	x_{t+1} & = & 1 - a~ x_t^2 + y_t + u_t, \nonumber \\
	y_{t+1} & = & b~ x_t.
	\end{eqnarray}
	
	Here $u_t = u(x_t,y_t)$ and $10^3$ data points are used to to train the Gaussian Process with parameters, $l=9$ and $q=64$.

	\begin{figure}[h]
		\centerline{
			\includegraphics[width=0.7\textwidth]{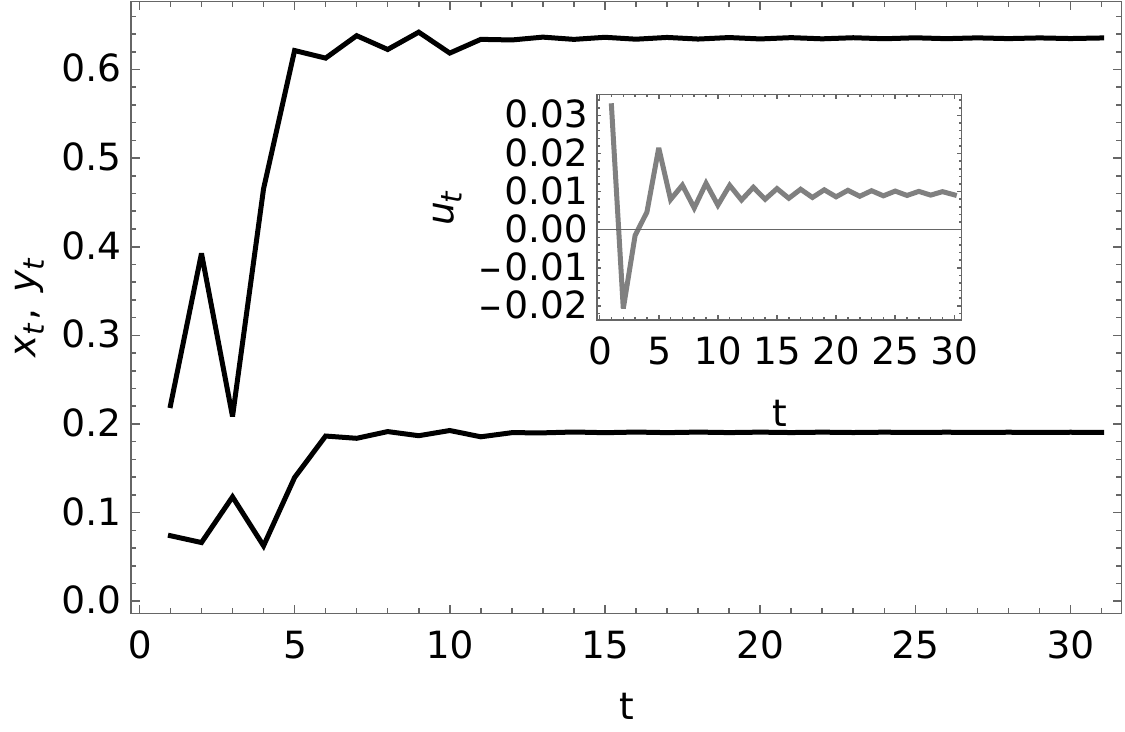}
		}
		\caption{\small Henon map. Evolution of the controlled system. The inset shows the applied perturbations. In both cases, the results shown were averaged for $100$ initial random conditions.}
		\label{Fig3}
	\end{figure}
	
	\subsection{Lorenz system}
	The Lorenz system is the typical example of the continuous chaotic dynamical system represented by the set of ordinary differential equations. With parameters $\sigma=10$, $\rho=28$ and $\beta=8/3$, it has unstable fixed points at $(\pm \sqrt{\beta (\rho-1)}, \pm \sqrt{\beta (\rho-1)}, \rho - 1)$.
	
	The controlled version of the system, in this case is:
	\begin{eqnarray}
	\dot x(t) & = & \sigma (y(t) - x(t)), \nonumber \\
	\dot y(t) & = & -x(t) z(t) + \rho x(t) - y(t), \nonumber \\
	\dot z(t) & = & x(t) y(t)- \beta z(t)  + u_t . \nonumber
	\end{eqnarray}
	
	Here $u(t) = u(x(t),y(t),z(t))$ and $1.2 \times 10^4$ data points are used to to train the Random Forest Regression,  with parameters, $l=10$ and $q=128$.
	
	\begin{figure}[h]
		\centerline{
			\includegraphics[width=0.7\textwidth]{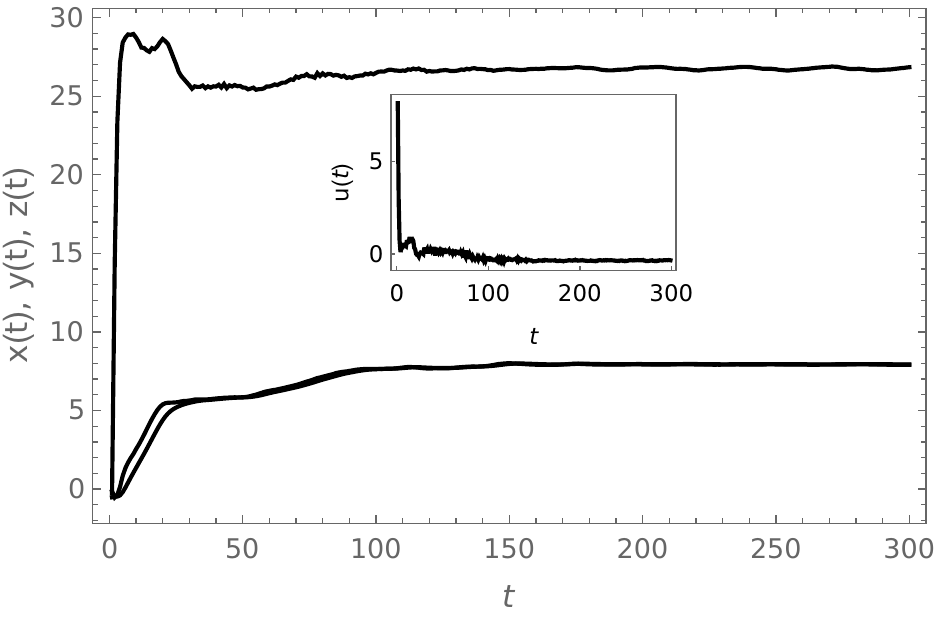}
		}
		\caption{\small Lorenz system. Evolution of the controlled system. The inset shows the applied perturbations.  In both cases, the results shown were averaged for $100$ initial random conditions.}
		\label{Fig4}
	\end{figure}
	
	\subsection{A coupled map network}
	The systems composed of parts that have some degree of autonomy and  interact, are frequent in innumerable situations and in areas from basic  to  applied science, so they must be an obligatory example of control object. Here, without loss of generality we use the logistic map, $f(s^i_{t}) = r s^i_t (1-s^i_t)$, as the nonlinear local dynamic to construct a generic complex dynamical network  consisting of $M$ identical nodes, 
	
	\begin{equation}
	s^i_{t+1} = f(s^i_{t}) -\frac{\epsilon}{k_i}\sum_{j=1}^{M} L_{ij} f(s^j_{t}),
	\label{CMN}
	\end{equation}
	
	\noindent
	where $L=(L_{ij})_{ij=1}^M$ is the usual Laplacian matrix with diagonal entries $L_{ii}=k_i$ and $k_i$ the out degree of the node $i$. Thus, the network (\ref{CMN}) can be represented as
	
	\begin{equation}
	{\bf s}_{t+1} = {\bf F}({\bf s}_{t})={\bf f}({\bf s}_{t}) - \epsilon~{\bf C}~{\bf f}({\bf s}_{t}),
	\label{CMN-M}
	\end{equation}
	
	\noindent
	where ${\bf s}_t$ are vectors, ${\bf f}({\bf s}_t) = (f(s^1_{t}), f(s^2_{t}),\cdots, f(s^M_{t}))$ a vector valued function, ${\bf C} = (L_{ij}/k_i))_{ij=1}^N$ is a matrix with real eigenvalues\cite{Xiang} and $\epsilon$ a real parameter.  Thus, the network can be represented as  ${\bf s}_{t+1} = {\bf A} {\bf f}({\bf s}_{t})$, with ${\bf A} = {\bf I}-\epsilon ~{\bf C}$. 
	
	This system have a unstable and homogeneous fixed point given by  ${\bf s}^*= (s_1^*, s_2^*, \dots, s_N^*)$. In this example $M=5$, the coupling parameter $\epsilon=0.1$ and
	
	\begin{equation}
	{\bf A} =
	\left(
	\begin{array}{cccccc}
	0.9 & 0. & 0.0333 & 0.0333 & 0.0333 & 0. \\
	0. & 0.9 & 0.025 & 0.025 & 0.025 & 0.025 \\
	0.0333 & 0.0333 & 0.9 & 0. & 0.0333 & 0. \\
	0.0333 & 0.0333 & 0. & 0.9 & 0.0333 & 0. \\
	0.02 & 0.02 & 0.02 & 0.02 & 0.9 & 0.02 \\
	0. & 0.05 & 0. & 0. & 0.05 & 0.9 \\
	\end{array}
	\right).
	\end{equation}
	
	\vspace{0.5 cm}
	
	\begin{figure}[h]
		\centerline{
			\includegraphics[width=0.5 \textwidth]{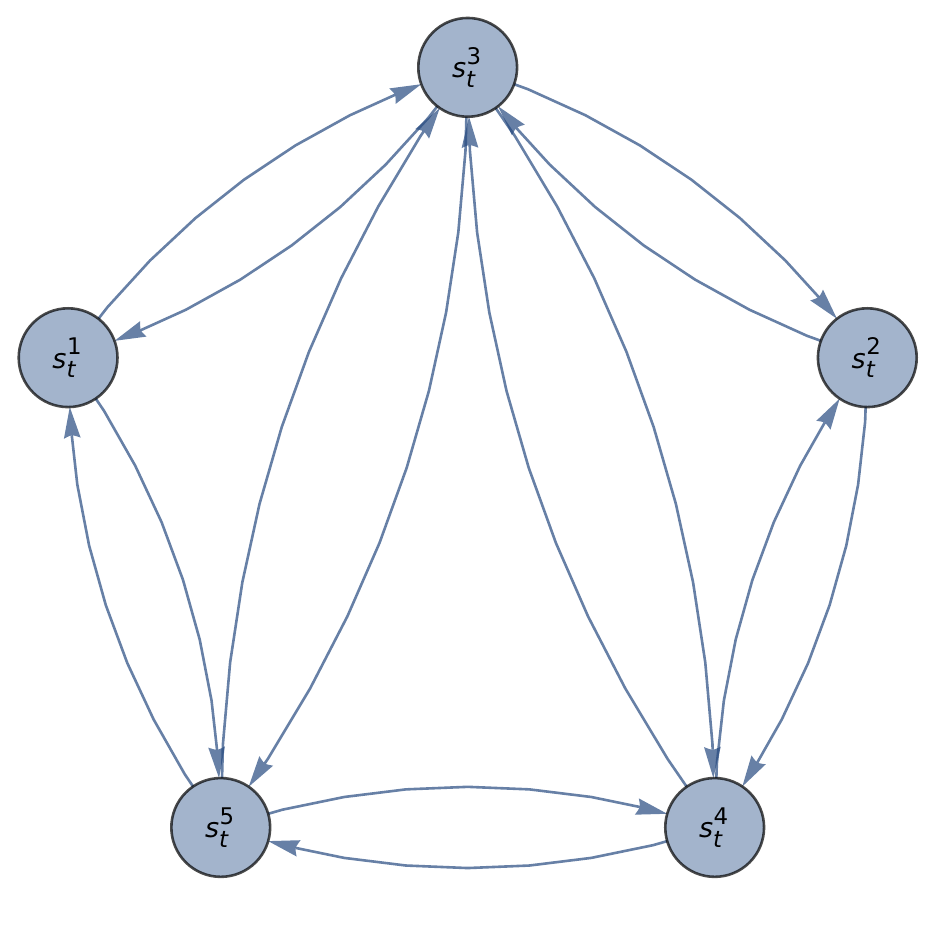}
		}
		\caption{\small Coupled chaotic maps network.}
		\label{Fig5}
	\end{figure}
	
	Although the symbolic dynamics of this network can be constructed in a way similar to the one-dimensional case \cite{Pethel}, here we will only present results for the data-dependent control strategy and although it is outside the our objectives the optimization of control sites we believe that this idea can be adapted to other contexts like the presented in \cite{Cornelius}. 
	
	The following figure shows evolution of the controlled system a a function of the time.
	
	\begin{figure}[h]
		\centerline{
			\includegraphics[width=0.7 \textwidth]{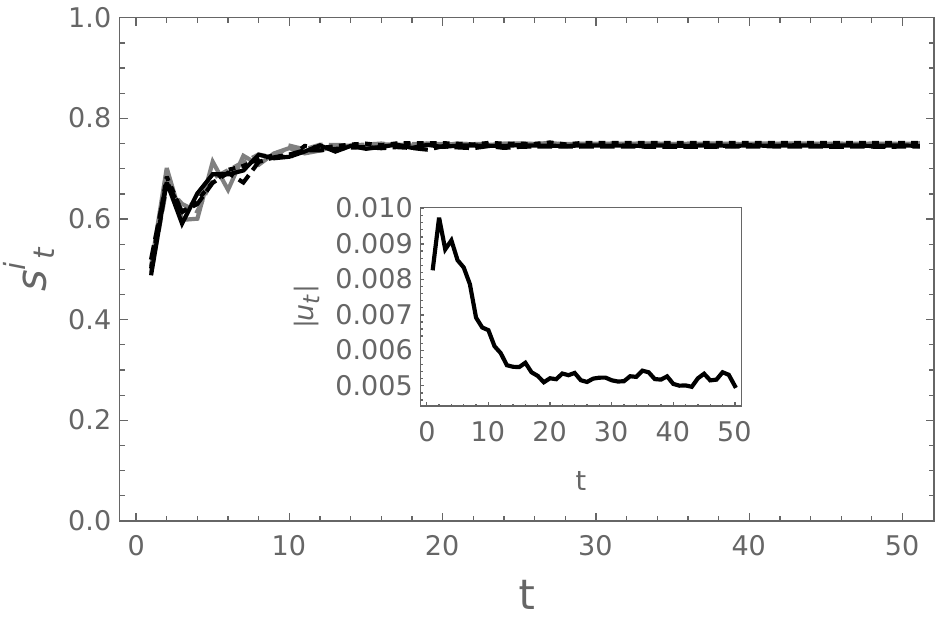}
		}
		\caption{Coupled map network. Evolution of the controlled system. Continuous black, dashed black, dotted black, continuous gray and dashed gray lines, shows the evolution of the five nodes, averaged for $100$ initial random conditions. The inset shows the applied perturbations.}
		\label{Fig6}
	\end{figure}
	
	In this example, the controlled system is written as:
	
	\begin{equation}
	s^i_{t+1} = f(s^i_{t}) -\frac{\epsilon}{k_i}\sum_{j=1}^{M} L_{ij} f(s^j_{t}) +  u^i_{t},
	\label{CMN-c}
	\end{equation}
	
	\noindent
	where $u^i_t=u^i({\bf s}_t)$.
	
	Figures \ref{Fig5} and \ref{Fig6} shows the structure of the network used as example in this article and the evolution of the controlled network, respectively. The control functions $u^i_t$ was generated by training the machine learning regression with $2.5 \times 10^5$ data points with parameters $l=7$ and $q=8 \times 10^3$.
	
	In spite of  the results shown in Figure \ref{Fig6}, refer to a network with the topology of the network in Figure \ref{Fig5}, the performance of the control scheme is similar for many networks with the same number of units. Numerical experiments suggest that, as this number increases the amount of data needed to obtain good approximations of $u$ quickly grows so it would be necessary, in the case of large networks, to combine the control strategy with a numerical methodology to approximate homoclinic orbits. This is out of the scope of this work and it is currently under investigation. 
	
	\section{Final remarks}  
	A  methodology for control of complex systems based on the estimation of the control function, from the observation of the system and using a machine learning technique, was presented.  Although in this particular case we have used the regression method based on Gaussian processes or Random Forest, the methodology allows us to use the regression technique that best suits the particular problem.
	
	Despite the existence of other control methods that use artificial intelligence strategi
	es, see for example \cite{Moe} and references therein, th
	e proposed method is, as far as we know,  novel in the sense that it directly approximate a nonlinear control signal from an rough estimate of homoclinic points and does so with very little computational effort.
	In addition, results suggest that the Proposed technique could be useful in the design of control or synchronization methodologies of complex systems like as \cite{Acosta}.
	
	Several aspects, although outside the scope of this work, remain as pending tasks. Among these aspects stand out: the optimization of the parameters ($q$, $k$ and $l$), selection of the sites to disturb or the design of a general criterion for the selection of regression strategy according to the number of data and dimension of the system.
	
	Finally, the aim of the work is to draw attention to two ideas: i) it is possible to convert the control target into a set of targets, which in some cases would simplify the control task, ii) it is possible to approximate the control $u(s_t)$ (not the system) using machine learning techniques. The results suggest that it is  possible to implement both ideas into a simple control algorithm.

\end{document}